\documentclass[letterpaper]{article} 
\ifdefined\aaaianonymous
    \usepackage[submission]{aaai2026}  
\else
    \usepackage{aaai2026}              
\fi
\usepackage{times}  
\usepackage{helvet}  
\usepackage{courier}  
\usepackage[hyphens]{url}  
\usepackage{graphicx} 
\usepackage{natbib}  
\usepackage{caption} 
\usepackage{algorithm}
\usepackage{algorithmic}
\usepackage{amsmath}
\usepackage{booktabs}
\usepackage{multirow}
\usepackage{xcolor}
\usepackage{subfigure}
\usepackage{times}
\usepackage{helvet}
\usepackage{courier}
\usepackage{xcolor}

\usepackage{newfloat}
\usepackage{listings}
\DeclareCaptionStyle{ruled}{labelfont=normalfont,labelsep=colon,strut=off} 
\floatstyle{ruled}
\newfloat{listing}{tb}{lst}{}
\floatname{listing}{Listing}
\newcommand{\name}{{\textsc{CoRec}}}
 \nocopyright 

\title{Token-Controlled Re-ranking for Sequential
Recommendation via LLMs}
\author{
    Wenxi Dai\textsuperscript{\dag},
    Wujiang Xu,
    Pinhuan Wang,
    Dimitris N. Metaxas\\
}
\affiliations{
    \textsuperscript{}Rutgers University\\
    wd232@scarletmail.rutgers.edu
    
}

\begin{document}

\maketitle

\begin{abstract}
The widespread adoption of Large Language Models (LLMs) as re-rankers is shifting recommender systems towards a user-centric paradigm. However, a significant gap remains: current re-rankers often lack mechanisms for fine-grained user control. They struggle to balance inherent user preferences with multiple attribute-based constraints, often resorting to simplistic hard filtering that can excessively narrow the recommendation pool and yield suboptimal results. This limitation leaves users as passive recipients rather than active collaborators in the recommendation process. To bridge this gap, we propose {\name}, a novel token-augmented re-ranking framework that incorporates specific user requirements in co-creating the recommendation outcome. {\name} empowers users to steer re-ranking results with precise and flexible control via explicit, attribute-based signals. The framework learns to balance these commands against latent preferences, yielding rankings that adhere to user instructions without sacrificing personalization. Experiments show that {\name}: (1) exceeds state-of-the-art baselines on standard recommendation effectiveness and (2) demonstrates superior adherence to specific attribute requirements, proving that {\name} enables fine-grained and predictable manipulation of the rankings.

\end{abstract}

\section{Introduction}
The advancement of LLMs has revolutionized tasks grounded in natural language, offering new frontiers for recommender systems. Compared to traditional approaches, LLMs significantly enhance the accessibility of recommender systems. Specifically, for next-item prediction tasks, we can leverage LLMs' superior natural language understanding capabilities to generate context-sensitive recommendations based on users' textual history \cite{chao2024makelargelanguagemodel,xu2025slmrecdistillinglargelanguage}. However, this paradigm faces critical limitations: users have limited direct control over recommendation outcomes, and there exists inherent latency in effectively incorporating user feedback. Consequently, users may experience a disconnect between algorithmic suggestions and their authentic preferences, therefore compromising user experience \cite{Aguirre2015UnravelingTP}. 

\begin{figure}[ht]
    \centering
    \subfigure[]{
        \includegraphics[width=0.3\textwidth]{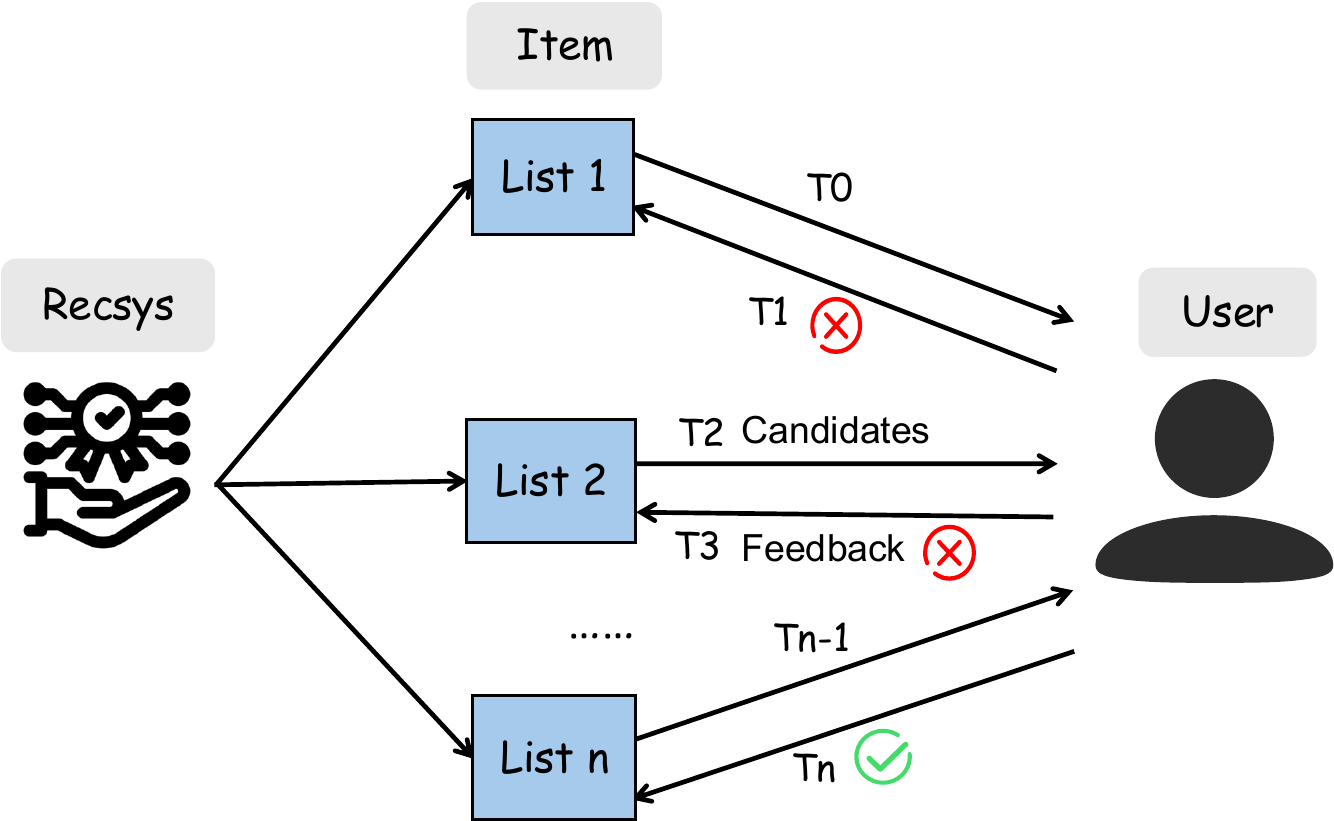}
        \label{fig:a}
    }
    \subfigure[]{
        \includegraphics[width=0.3\textwidth]{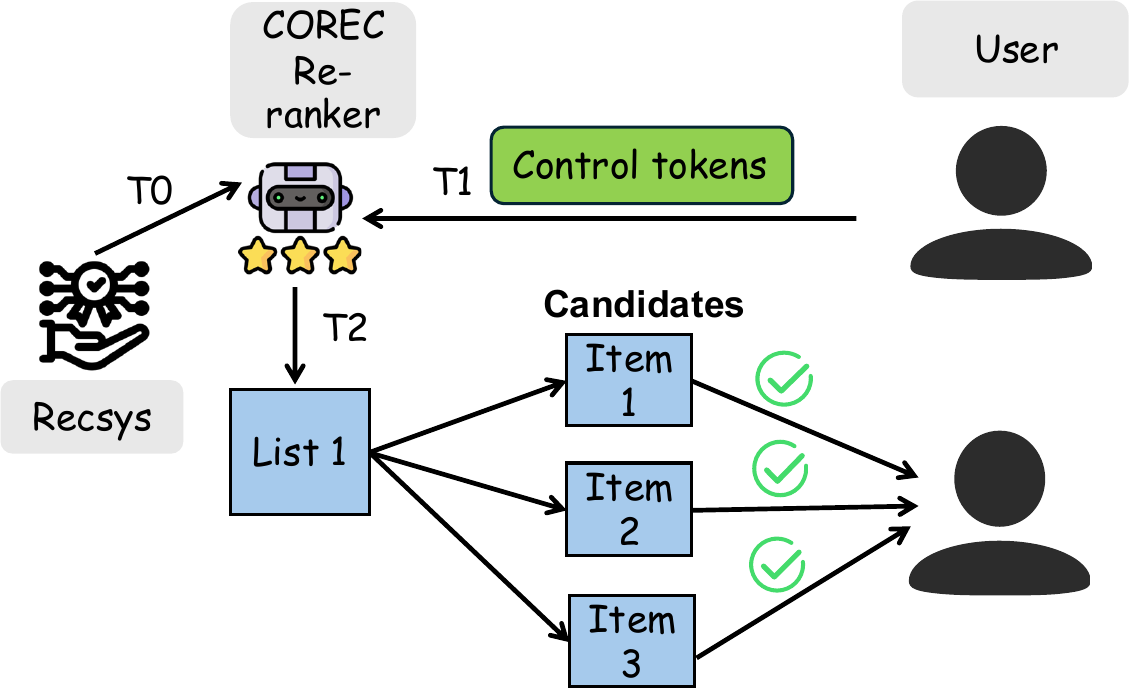}
        \label{fig:b}
    }
    \caption{(a) Previous methods in recommender systems have largely focused on inferring potential preferences. This paradigm places users in a passive position. (b) In contrast, {\name} allows users to regulate recommendation results according to fine-grained attributes.}
    \label{intro_comparison}
\end{figure}
These limitations highlight a growing need for more controllable and responsive systems. The emergence of intelligent agents now enables systems to leverage explicit user feedback to continuously enhance model performance and adaptability \cite{gao2025llm4rerankllmbasedautorerankingframework,huang2024recommenderaiagentintegrating, xu2025iagentllmagentshield}. This empowers users to proactively express their intent, shifting them from passive recipients to active initiators who set tasks and engage in interaction. Such proactive expressions of intent, however, often involve specific preferences and constraints on item attributes \cite{wu2023surveyLLM4Rec}. 

To effectively address this, LLM-based recommender systems are required to perform specific planning and trade-offs at the item attribute level \cite{zhang2023recommendationinstructionfollowinglarge, gao2025llm4rerankllmbasedautorerankingframework, chao2024makelargelanguagemodel}. However, in most current systems, mapping user intent to a set of attribute preferences remains an implicit part of a singular inference process. This introduces several challenges: Firstly, limited fine-grained user control is provided when generating recommendations. Current methods predominantly focus on coarse-grained adjustments, guiding the model towards high-level objectives such as diversity or fairness \cite{qu2024tokenreclearningtokenizeid,gao2025llm4rerankllmbasedautorerankingframework,chen-etal-2025-DLCRec}. While valuable for shaping overall recommendation behavior, these methods systematically function as broad guidance across the entire user base, rather than enabling precise, attribute-level control for individual users over their personalized outputs. Secondly, methods that do engage with item attributes typically resort to hard filtering \cite{Zhao_2024}. This binary "include-or-exclude" logic, while straightforward, can prove suboptimal in nuanced scenarios. When multiple desired attributes are introduced simultaneously, hard filtering tends to become overly restrictive, potentially resulting in empty recommendation sets. 

To bridge this gap, we propose {\name}, a novel token-augmented re-ranking framework for sequential recommendation. {\name} is designed to perform controlled, context-aware reasoning for recommendation results.

Our main contributions are summarized as follows:
\begin{itemize}
\item\textbf{Control Mechanism For Recommendation:} We propose and formulate the concept of \textbf{control tokens}, a novel mechanism to explicitly encode fine-grained, attribute-level user requirements into a structured LLM re-ranking process. These tokens function by disentangling users' requirements into fine-grained tokens that guide the model's re-ranking process. This creates a more structured and transparent reasoning path, enabling users to control the reasoning behind recommendation results, therefore enhancing the system's interpretability.

\item\textbf{Token-Augmented Recommendation Framework:} We design and implement {\name}, a comprehensive token-augmented re-ranking framework. {\name} integrates control token construction with textual user information, enabling flexible control schemes to augment recommendation results through user control. The framework is specifically optimized to balance user preferences with explicit, on-demand constraints. This approach delivers personalized recommendations tailored to each user's specific attribute requirements, moving beyond broad adjustments that target the entire user base.

\item\textbf{Empirical Evaluation:} We conduct extensive experiments using different combinations of control tokens to validate the effectiveness of our approach. Results show that {\name}: (1) maintains competitive recommendation performance and (2) surpasses existing baselines in adhering to fine-grained user constraints, as measured by newly introduced controllability metrics. These findings confirm that our framework successfully enhances user control while balancing recommendation quality.
\end{itemize}

\section{Related Works}

\subsection{LLM-based Re-ranker in Recommendation}
Modern recommender systems often involve processing large candidate pools, making direct LLM-based ranking computationally expensive. To leverage the capabilities of LLMs while maintaining efficiency, recent approaches have developed a two-stage re-ranking framework. This framework first employs lightweight methods in the retrieval stage to pre-filter candidate items, then applies LLM-based ranking in the second stage \cite{hou2024largelanguagemodelszeroshot,wu2023surveyLLM4Rec,yue2023llamarectwostagerecommendationusing, rathee2025guidingretrievalusingllmbased}, thereby avoiding the need to process the entire item pool. Within this paradigm, different approaches have emerged for leveraging LLMs in ranking tasks. One line of work employs generative ranking, where LLMs directly generate ranked candidate item sequences, similar to standard text-generation paradigms in generative recommendation \cite{gao2025llm4rerankllmbasedautorerankingframework,wu2023surveyLLM4Rec, li2024largelanguagemodelsgenerative}. However, this approach can incur high computational costs due to the need to generate complete recommendation sequences.

In contrast, scoring-based methods take a more efficient approach. These methods either assign scores to candidates directly \cite{chao2024makelargelanguagemodel, qin-etal-2024-large, xin2025llmsbetterrecommendersnatural, liu2025corankingcollaborativerankingsmall} or generate index tokens only. The probabilities of these tokens are then extracted from the LLM's output logits and converted into ranking scores for the candidates\cite{yin2024unleashllmspotentialrecommendation, yue2023llamarectwostagerecommendationusing, gangi-reddy-etal-2024-first, qu2024tokenreclearningtokenizeid,wang-etal-2025-realm}. Given the reduced computational cost and effective ranking quality, this approach shows promising potential for practical applications.

\subsection{Guiding Generative Processes in Recommendation}
Recent studies have demonstrated that inserting learnable tokens to control the reasoning steps can enhance both the training and inference stages of LLMs. These discrete and semantically meaningful tokens enable various forms of generation control in LLMs, such as allocating additional computation time before generation \cite{goyal2024thinkspeaktraininglanguage} or guiding the model’s chain-of-thought reasoning process \cite{zelikman2024quietstarlanguagemodelsteach, wang2024guidinglanguagemodelreasoning,jin2024impact,jin-etal-2025-disentangling-memory}. 

This fine-grained, token-based control creates structured reasoning paths that allow users to influence recommendation outcomes directly. While traditional recommender systems often leave users in passive roles, recent approaches have explored various mechanisms to enhance user agency in recommendation. Some methods enforce strict attribute constraints through specialized architectural designs \cite{Zhao_2024, gao2025futureconditionedrecommendationsmultiobjectivecontrollable, zhang2023recommendationinstructionfollowinglarge}, which primarily focus on promoting specific objectives rather than providing flexible control. Other approaches aim to enhance recommendation diversity and overcome passive user acceptance through prompt-based methods that decompose recommendation tasks—either by categorizing into different types \cite{chen-etal-2025-DLCRec, lu-etal-2024-aligning} or through multi-step reasoning \cite{gao2025llm4rerankllmbasedautorerankingframework, fang2025reason4reclargelanguagemodels}. These prompting strategies typically bias recommendations toward broad preferences (e.g., diversity or fairness) but do not aim for precise, quantifiable control over specific item attributes. In contrast, our work focuses on fine-grained control of existing item attributes through a re-ranking model that balances flexibility with precise control. We provide a more detailed review of related works in Appendix A.

\section{Methodology}
In this chapter, we first formulate the problem of controlled LLM-based recommendation, establishing the theoretical foundation for our method. We then present the {\name} framework workflow, as illustrated in Figure~\ref{pipeline}, detailing how our method enables precise injection of control tokens into the recommendation model. Finally, we introduce a novel ranking loss objective that enables the model to generate rankings that are both aligned with users' historical preferences and responsive to structured control token inputs.

\subsection{Problem Formulation}
We consider a standard recommendation setup where $\mathcal{U}$ denotes the set of users and \(\mathcal{I}\) the set of items. For a user \(u \in \mathcal{U}\), we represent their interaction history as a sequence \(\mathcal{H}_u = [i_1, i_2, \dots, i_t]\), where each \(i_k \in \mathcal{I}\) denotes an item the user previously interacted with. We also define a set of control attributes \(N_C\) (e.g., price, brand), therefore forming a sequence \(N_C = \{c_1, \ldots, c_M\}.\)

In \textbf{sequential recommendation}, the goal is to learn a function \(f_\theta\) that estimates a conditional probability distribution over items for the next interaction \cite{sun2019bert4recsequentialrecommendationbidirectional}. Given a user’s history \(\mathcal{H}_u\), the model recommends the next item \(\hat{i}\) according to the following formulation:
\begin{equation}
\hat{i} = \arg\max_{i \in \mathcal{I}} f_\theta(\mathcal{H}_u, i)
\label{sequential}
\end{equation}

Building on Eq.\ref{sequential}, we extend our method for \textbf{LLM-based sequential recommendation}. We adopt a two-stage approach in which a base retriever returns a user-specific candidate set $I_u^{\text{cand}}$. Then the LLM \textbf{re-ranks} these candidates conditioned on the user's interaction history and control tokens. In our approach, the user interaction history \(\mathcal{H}_u\) is represented as a textual sequence.
Specifically, an LLM parameterized by \(\theta\) models the probability of each candidate item and selects the one with the highest probability:

\begin{equation}
\hat{i} = \arg \max_{i \in I_u^{cand}} p_\theta(w_i \mid \mathcal{H}_u, N_C)
\end{equation}
where \(w_i\) represents the next item selected from the candidate set $I_u^{\text{cand}}$, and the model jointly considers user interaction history and control token constraints.

\begin{figure}
    \centerline{\includegraphics[width=0.75\linewidth]{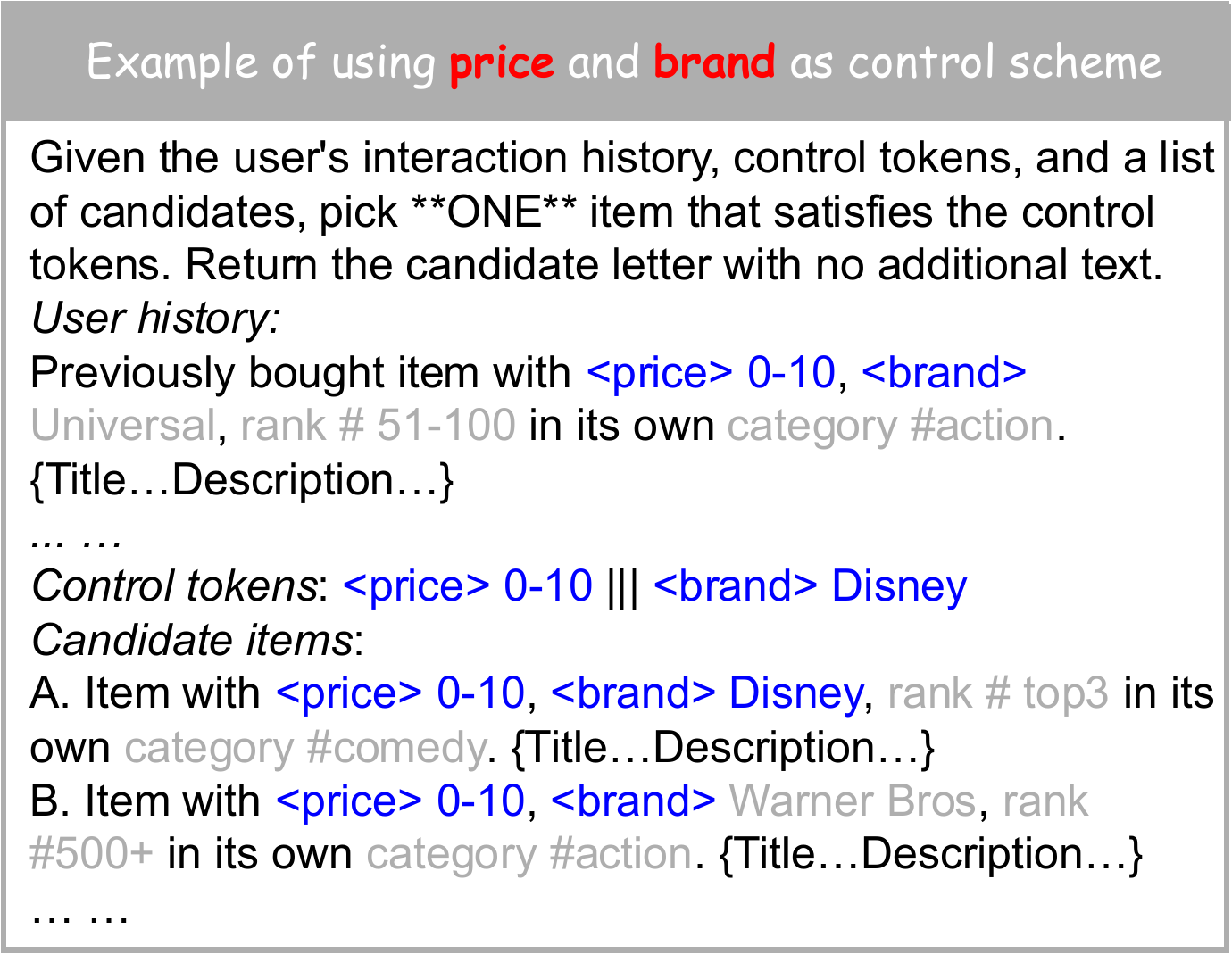}}
    \caption{Illustration of our input. We use blue to mark components aligned with the control scheme: the scheme itself and matching control tokens. Gray is used for non-matching token elements and uncontrolled metadata.
    }
    \label{prompt_total}
\end{figure}
\subsection{{\name} Pipeline}
\begin{figure*}
    \centerline{\includegraphics[width=1\linewidth]{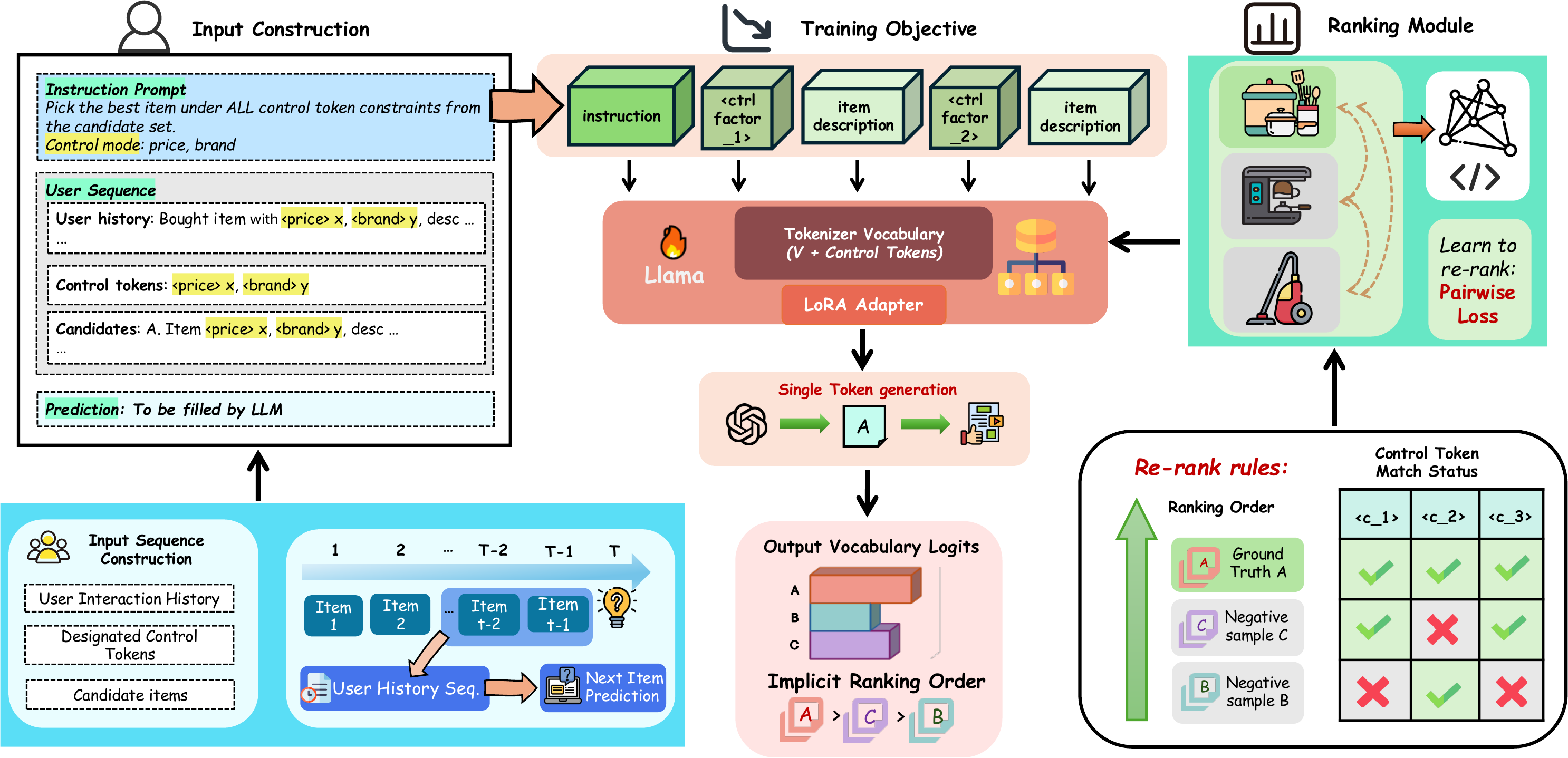}}
    \caption{Pipeline of {\name}. In the retrieval stage, we generate candidate items for each window. During the LLM re-ranking stage, the input sequence is constructed, injecting control tokens, and the LLM is then trained based on the ranking module.
    }
    \label{pipeline}
\end{figure*}

In this section, we introduce the {\name} pipeline, which follows a two-stage workflow similar to industrial recommender systems. We present the methodology following the sequential steps outlined in Figure~\ref{pipeline}.

\subsubsection{Item Retrieval Stage.}
We adopt the classic SASRec \cite{kang2018sasrec} as our lightweight sequential recommendation method for the item retrieval stage. SASRec leverages self-attention over sequences of unique item IDs to capture users' interests. To generate candidates for re-ranking, we employ a sliding window approach over the user's historical interactions, as illustrated in the bottom left of Figure~\ref{pipeline}. From this stage, we pre-filter the top-\(K\) candidates to serve as the candidates for subsequent re-ranking. The candidate set, along with the user's historical sequence, will be used to construct the input for the LLM re-ranker. 

\subsubsection{Input Construction.}

Our input representation framework integrates structured control signals, namely \textbf{control tokens}, with textual user information. These tokens are designed to embed attribute-level signals directly into the input sequence, enabling controlled and planned recommendations.

In the context of sequential recommendation, we apply this framework to construct an attribute-guided user interaction sequence. When a user defines a control scheme, for instance, using \textbf{price} and \textbf{brand}, they're essentially employing \textbf{two} distinct types of control tokens. For instance, as in Figure~\ref{prompt_total}, consider a user who wishes to prioritize purchasing videos in the lower price range of \$0-\$10, and children's entertainment from the Disney brand, while having no specific requirements for the video category. Consequently, the control tokens for this user are \texttt{<price>0-10} and \texttt{<brand>Disney}. 

To implement this, we systematically replace the corresponding natural language information in the interaction sequence with our standardized control tokens. Continuing the example, the price attribute of a candidate item A, originally \$6, is categorized into the \$0-\$10 bucket and then converted into its control token form. Likewise, the item’s brand is mapped to its corresponding control token. Other metadata not specified in the current control scheme, such as rank and category, are preserved in their original form and serve as supplementary information in the prompt. Therefore, the final input for the model is a hybrid sequence comprising both the structured control tokens and the original textual information. Finally, following the workflow in Figure~\ref{pipeline}, this input is paired with an instruction prompt that constrains the output format. The complete sequence is later used to fine-tune the LLM for the next item prediction task. Notably, for the fine-tuning stage, \texttt{<price>} and \texttt{<brand>} are incorporated into the tokenizer's vocabulary as special tokens, as shown in the tokenizer vocabulary expansion in Figure~\ref{pipeline}.

\subsubsection{Training Objective and Ranking Module.}
Our fine-tuning objective employs a learning-to-rank loss. Specifically, we utilize a pairwise RankNet loss \cite{Ranknet} to optimize the relative ordering of candidate items. From the LLM head, we extract logits corresponding to index letters (A to Z), where each letter represents a unique candidate item. The resulting probability distribution serves as the ranking scores \(s_i\) for these candidates. It's important to note that only these candidate index letters participate in the loss computation, as our task focuses on ranking rather than traditional language modeling.

We assign control scores, denoted as \(r_i\), to each candidate item using a carefully designed scoring rule. This rule simulates a robust user modeling process and is provided to the LLM to encourage specific control over the re-ranking output. 
Here, ground truth (we mark as GT) is the user's actual next interacted item in our recommendation task.

Our scoring mechanism operates as follows: Let $N_C$ be the number of control tokens for the current sequence. For candidate item $i$, let $s_{i,k}\in\{0,1\}$ indicate whether control token $k$ is satisfied ($k=1,\dots,N_C$), and define \(m_i=\sum_{k=1}^{N_C} s_{i,k}\) as the cumulative count of satisfied tokens. Let $g$ index the GT item if present in the candidate set; otherwise, no item is marked GT. We assign control scores as:
\begin{equation}
r_i = m_i + \mathbf{1}[i=g].
\end{equation}
Thus, a ground truth item that satisfies all control tokens receives $N_C+1$ points; a ground truth item that misses some tokens receives $m_g+1$; non-ground truth items receive $m_i$; and when the ground truth item is absent from candidates, ranking reduces to control-token adherence ($r_i=m_i$).

For instance, following on the previous Figure \ref{prompt_total}, as control tokens specify \texttt{<price>0-10} and \texttt{<brand>Disney}, suppose the candidate B movie is the ground truth item. It would receive the same control ranking weight as candidate A, even if it doesn't perfectly align with all specified criteria.

Given \(K\) candidate items, we pre-process and assign control scores based on the given tokens. We then compare all pairs of these items where their control scores are unequal. Our comparison is based on the principle that for any two items, if one has a higher control score \(r_i\), its model-predicted re-rank score \(s_i\) should also be raised with higher weights than that of the item with a lower control score, which can be formally stated as:
\begin{equation}
\forall\, i,j\in\{1,\dots,K\},\;
r_i > r_j \;\Longrightarrow\; s_i > s_j,
\end{equation}
We define the set of all valid ordered pairs \(\mathcal{P}\) among the \(K\) candidates as:
\(
\mathcal{P} = \left\{ (i, j) \;\middle|\; 1 \leq i, j \leq K,\; r_i > r_j \right\}
\)
. We minimize the final pairwise loss as follows:
\begin{equation}
\mathcal{L}_{\text{rank}} = \frac{1}{|\mathcal{P}|} \sum_{(i, j) \in \mathcal{P}} -\log \sigma(s_i - s_j)
\end{equation}
\section{Experiments}

In this section, we conducted extensive experiments to demonstrate the effectiveness of {\name}, answering the following research questions \textbf{(RQs)}.
\begin{itemize}
\item\textbf{RQ1:} How's the overall performance of {\name} compared to various baselines across different datasets?
\item\textbf{RQ2:} Does the variety and quantity of control tokens influence recommendation and control effectiveness?
\item\textbf{RQ3:} Is the proposed {\name} model capable of precisely controlling recommendation outputs according to user-defined item attribute aspects?
\item\textbf{RQ4:} Can {\name} model balance user control and recommendation performance better than filtering?
\end{itemize}
\subsection{Implementing Details.}
\subsubsection{Training.}
We employ Llama3-8B as our foundation model and apply LoRA \cite{LoRA} for parameter-efficient fine-tuning. The model was trained with a pairwise ranking loss objective using the AdamW optimizer \cite{AdamW}. We used a learning rate of [1e-4, 3e-5, 7e-6] with a cosine scheduler and a 2\% warmup period. Fine-tuning was performed for each control-token combination, including configurations with one, two, or three control tokens in our setup on four RTX A5000 GPUs (24 GB), with 3 epochs, accelerated by DeepSpeed \cite{deepspeed} using bfloat16 mixed-precision and gradient checkpointing. We provide further details in Appendix B.

\begin{table}[t]
\centering
\caption{Statistics of preprocessed datasets used to evaluate {\name}: \(|\mathcal{U}|\), \(|\mathcal{I}|\), and \(|\mathcal{E}|\) represent the number of users, items, and interactions, respectively. }
\label{dataset stats}
\small           
\setlength{\tabcolsep}{4pt}  
\begin{tabular}{cccccc}
\hline
\textbf{Dataset} & $|\mathcal{U}|$ & $|\mathcal{I}|$ & \textbf{$|\mathcal{E}|$} & \textbf{Density}\\
\hline
\textbf{Home} & 5,555 & 70,585 & 249,965 & 0.0638\%\\
\textbf{Electronics}      & 4,783 & 49,234  & 211,343     & 0.0897\%  \\
\textbf{Beauty}       & 624  & 1,248  & 4,239      & 0.5443\% \\
\hline
\end{tabular}
\end{table}

\subsubsection{Datasets.} 
To validate the effectiveness of {\name}'s control mechanism, we conduct experiments using control tokens derived from specific item attributes. Implementing different combinations of these tokens requires rich and varied metadata. Therefore, we utilize data from Home and Kitchen, Electronics, and All Beauty categories within the extensive Amazon 2018 dataset \cite{ni-etal-2019-amazon2018}. These datasets provide comprehensive textual information, including item titles and descriptions, as well as item attributes such as price, brand, and ranking, which serve as different aspects of control tokens in our experiments. The statistical analysis of our datasets is illustrated in Table~\ref{dataset stats}.

We first pre-filter the data to ensure all control tokens contain non-blank values, namely price, rank, and brand for Home and Electronics datasets and category for Beauty dataset. Following the methodology in previous methods \cite{kang2018sasrec, zhang2023recommendationinstructionfollowinglarge}, we partition each user's interaction sequence into three sets: the most recent 10\% of interactions are used for testing, the preceding 10\% for validation, and the remainder for training. The input sequence length was configured to 6 using a sliding window sampling. For the re-ranking evaluation task, we constructed a compact candidate set for each target item. 

For the Beauty dataset, we employ augmented control tokens since it lacks sufficient native metadata. Specifically, we create product category tags using the GPT-4 API \cite{openai2024gpt4technicalreport} to generate tags that describe product characteristics. To prevent data leakage, all control token settings are defined exclusively using the training split. This includes the augmentation of category tags for the Beauty dataset and the distribution-based bucketing for price and rank tokens. The bucket boundaries for price and rank are established based solely on training data statistics, ensuring no information from validation or test sets influences the pre-processing pipeline. More detailed data pre-processing, data augmentation, and filtering processes are provided in Appendix B.

\begin{figure}[ht]
    \centering
    \subfigure[]{
        \includegraphics[width=0.165\textwidth]{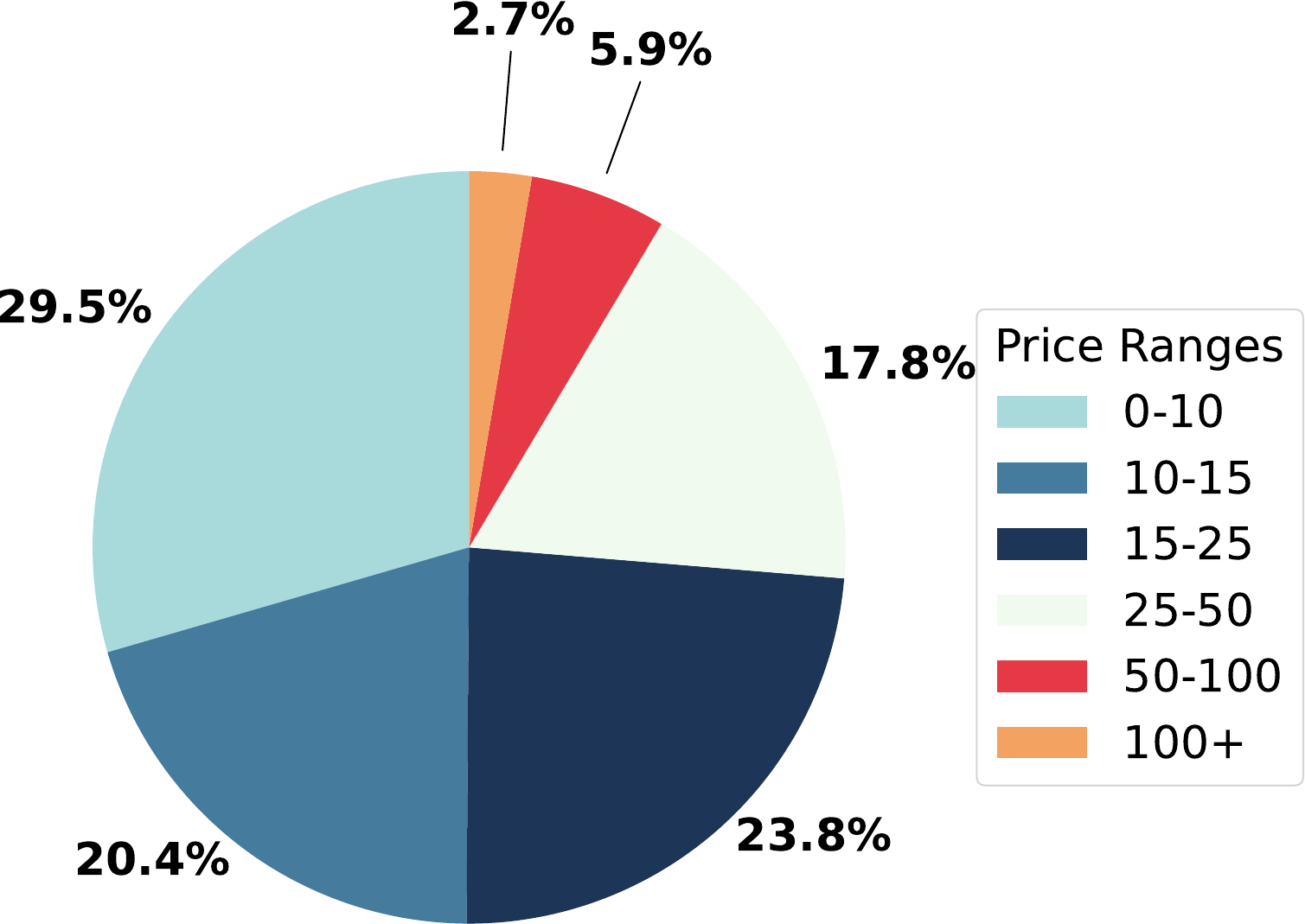}
        \label{fig:a}
    }
    \subfigure[]{
        \includegraphics[width=0.28\textwidth]{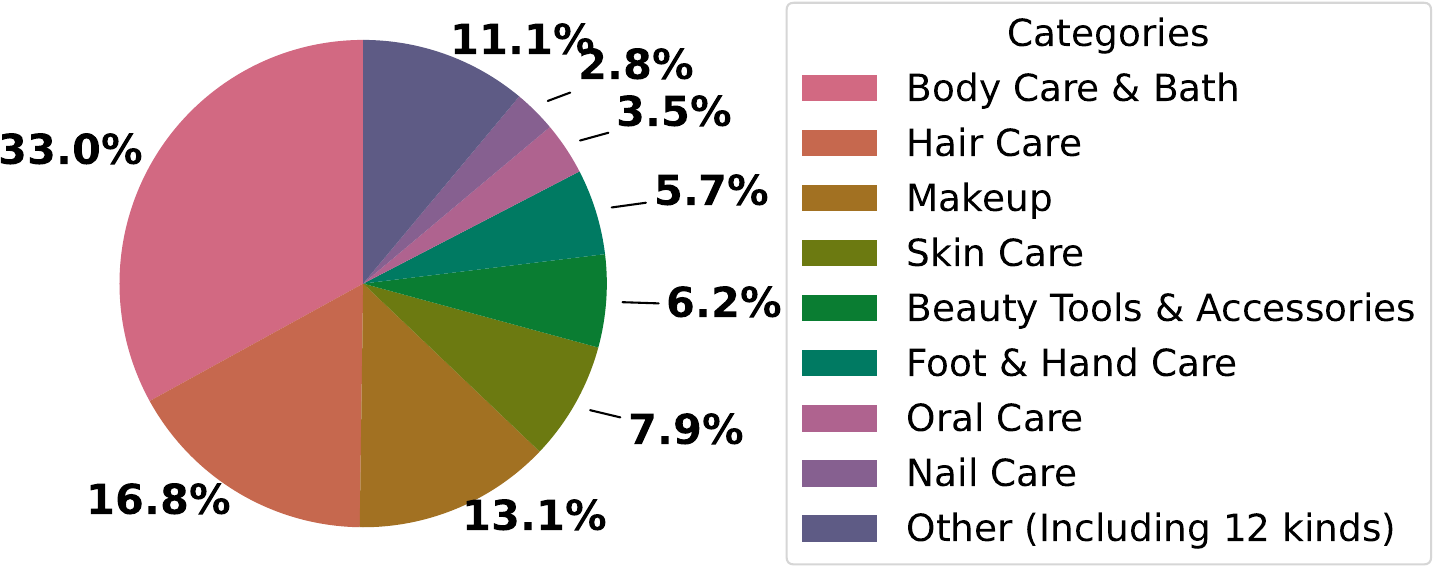}
        \label{fig:b}
    }
    \caption{(a) Distribution of \texttt{<price>} in Home training set. (b) Distribution of \texttt{<category>} in Beauty training set.}
    \label{distribution_comparison}
\end{figure}

The distribution of the control token types discussed above is shown in Figure \ref{distribution_comparison}. For each constructed user sequence, we select the control tokens by randomly sampling from the available attributes of the six candidate items, in accordance with our designated control scheme.

\subsubsection{Evaluation.}
For Home and Electronics datasets, we design three control schemes with varying numbers of control tokens: (1) a \textbf{single-token} scheme controlling price, (2) a \textbf{two-token} scheme controlling price and rank, and (3) a \textbf{three-token} scheme controlling price, rank, and brand. For the Beauty dataset, we employ category tags obtained through data augmentation as a single control token. 

We evaluate the overall performance of our framework using standard top-N metrics, including normalized discounted cumulative gain (NDCG@\{2,5\}) and hit rate (HR@\{3\}). To capture the nuanced preferences reflected in our control mechanism, we leverage control scores as graded relevance in NDCG computation, enabling a more fine-grained assessment of ranking quality. To ensure a comprehensive evaluation, we complement this with ground truth hit rate analysis using binary relevance, where only the actual ground truth item is considered relevant. This evaluation approach eliminates the constraints of control tokens and objectively assesses whether our method maintains advantages in ranking top items. Models with the highest validation hit rates will be evaluated on the test set. For consistent comparison across different approaches, all baseline methods are re-evaluated under identical control schemes.

In addition to traditional ranking metrics, we introduce two novel evaluation metrics to validate the effectiveness of our control implementation. Specifically, we design: 

\begin{enumerate}

 \item Control Precision (CP@K): 
 
 For a user \(u\) with a top-\(K\) recommendation list \(S_u\), we define Control Precision (CP@K) as the fraction of items in \(S_u\) whose relevance scores exceed threshold relevance \(t\):

\begin{equation}
\mathrm{CP@K}(u) = \frac{\left| \left\{ i \in S_u \mid \operatorname{rel}(i) \ge t \right\} \right|}{K}
\end{equation}
where  \(\operatorname{rel}(i)\) is the graded relevance of item \(i\), and \(t\) represents the maximum relevance that a controlled item can achieve using \(N_C\) kinds of control tokens, and \(t=N_C\).

 \item Control Depth (CD): 

Control Depth measures how early a control-compliant item appears in the ranked list \(S_u\). A lower CD indicates better compliance with control tokens. Formally:

\begin{equation}
\mathrm{CD}(u)=
\begin{cases}
\displaystyle\min_{i}\,\operatorname{rank}(i) & \text{if }\exists\, i\in S_u\text{ with }\operatorname{rel}_i \ge t,\\
K+1 & \text{otherwise}.
\end{cases}
\end{equation}
\end{enumerate}

\begin{table*}
\centering
\caption{Results on {\name}. We mark the best results for each scheme in bold and underline the second-best results.}
\label{COREC-results_1}
\small
\begin{tabular}{c|c|ccccc|ccccc}
\hline
\textbf{Control} & \multirow{2}{*}{\textbf{Model}} & \multicolumn{5}{c|}{\textbf{Home}} & \multicolumn{5}{c}{\textbf{Electronics}}\\
\textbf{token(s)} & & \textbf{N@2} & \textbf{N@5} & \textbf{H@3} & \textbf{CP@3} & \textbf{CD} & \textbf{N@2} & \textbf{N@5} & \textbf{H@3} & \textbf{CP@3} & \textbf{CD} \\
\hline
\multirow{8}{*}{\rotatebox{90}{\textbf{\textless price\textgreater}}} & SASRec & 0.2961 & 0.3120 & 0.1067 & 0.3035 & 2.9754 & 0.4219 & 0.4429 & 0.1015 & 0.4411 & 2.5156 \\
& BERT4Rec & 0.2653 & 0.2791 & 0.1024 & 0.2741 & 3.2891 & 0.3995 & 0.4137 & 0.0892 & 0.3712 & 2.8537 \\
& GRU4Rec & 0.2096 & 0.2122 & 0.0983 & 0.2328 & 3.4190 & 0.3340 & 0.3275 & 0.0679 & 0.3264 & 3.1992 \\
\cline{2-12}
& LLaRA & 0.5502 & 0.4270 & 0.1045 & 0.3865 & 2.0135 & 0.6502 & 0.5418 & 0.1259 & 0.4963 & 1.9618 \\
& LlamaRec & 0.2958 & 0.3112 & 0.0829 & 0.3029 & 2.9765 & 0.4241 & 0.4433 & 0.1121 & 0.4417 & 2.3098 \\

& TALLRec & \underline{0.6134} & \underline{0.4416} & \underline{0.1437} & \underline{0.4999} & \underline{1.8828} & \underline{0.7283} & \textbf{0.6634} & \underline{0.1473} & \underline{0.5430} & \underline{1.6796} \\
\cline{2-12}
& \textbf{{\name}} & \textbf{0.7882} & \textbf{0.5102} & \textbf{0.1641} & \textbf{0.5826} & \textbf{1.5502} & \textbf{0.8333} & \underline{0.6327} & \textbf{0.1507} & \textbf{0.7160} & \textbf{1.4830} \\
\hline
\multirow{8}{*}{\rotatebox{90}{\textbf{\textless price,rank\textgreater}}} & SASRec & 0.3126 & 0.3485 & 0.1061 & 0.1058 & 5.0181 & 0.4056 & 0.4361 & 0.1044 & 0.1641 & 4.4138\\
& BERT4Rec & 0.2934 & 0.3267 & 0.1264 & 0.1148 & 5.1649 & 0.3972 & 0.4145 & 0.0958 & 0.1842 & 4.6032 \\
& GRU4Rec & 0.2447 & 0.2607 & 0.0875 & 0.0474 & 5.2692 & 0.3408 & 0.3362 & 0.0712 & 0.0869 & 4.9654 \\
\cline{2-12}
& LLaRA & \underline{0.4109} & \underline{0.3264} & \underline{0.1253} & \underline{0.1108} & 4.4198 & 0.4527 & 0.3980 & 0.1465 & 0.1938 & 3.9746
 \\
& LlamaRec & 0.3101 & 0.3470 & 0.1138 & 0.1071 & 5.0207  & 0.4066 & 0.4358 & 0.1482 & 0.1647 & 4.1026\\
& TALLRec & 0.3381 & 0.3485 & 0.1046 & 0.1080 & \underline{4.4960} & \underline{0.5976} & \underline{0.5014} & \textbf{0.1584} & \underline{0.2382} & \underline{3.5742} \\
\cline{2-12}
& \textbf{{\name}} & \textbf{0.7123} & \textbf{0.5344} & \textbf{0.1689} & \textbf{0.2160} & \textbf{3.7880} & \textbf{0.7698} & \textbf{0.6127} & \underline{0.1558} & \textbf{0.3199} & \textbf{3.1383}\\
\hline
\multirow{8}{*}{\rotatebox{90}{\parbox{1.2cm}{\centering\textbf{\textless price,rank,\\ brand\textgreater}}}} & SASRec & 0.3354 & 0.3801 & 0.1064 & 0.0418 & 6.1173 & 0.4104 & 0.4560 & 0.1010 & 0.0457 & 6.0438\\
& BERT4Rec & 0.3127 & 0.3515 & 0.1184 & 0.0369 & 6.0852 & 0.3971 & 0.4288 & 0.1004 & 0.0468 & 6.1029 \\
& GRU4Rec &  0.2473 & 0.2568 & 0.0846 & 0.0173 & 5.9979& 0.3207 & 0.3339 & 0.0825 & 0.0231 & 5.9860 \\
\cline{2-12}
& LLaRA & 0.3754 & 0.4017 & \underline{0.1492} & \underline{0.0528} & 5.9967 & 0.4379 & 0.4853 & 0.1299 & 0.0493 & 5.9104 \\
& LlamaRec & 0.3350 & 0.3795 & 0.0976 & 0.0423 & 6.1118 & 0.4123 & 0.4574 & 
0.1347 & 0.0458 & 5.9873 \\
& TALLRec & \underline{0.4029} & \underline{0.4124} & 0.1465 & 0.0469 & \underline{5.9843} & \underline{0.4757} & \underline{0.4880} & \underline{0.1531} & \underline{0.0625} & \underline{5.5898} \\
\cline{2-12}
& \textbf{{\name}} & \textbf{0.7485} & \textbf{0.5799} & \textbf{0.1632} & \textbf{0.0846} & \textbf{5.4816} & \textbf{0.8010} & \textbf{0.6508} & \textbf{0.1605} & \textbf{0.0923} & \textbf{5.3442}\\
\hline
\end{tabular}
\end{table*}

\begin{table}[t]
\centering
\caption{Results on {\name} . Best results are marked in bold, and second-best results are underlined.}
\label{COREC-results_2}
\small
\setlength{\tabcolsep}{1mm}
\begin{tabular}{c|c|ccccc}
\hline
\textbf{Control} & \multirow{2}{*}{\textbf{Model}} & \multicolumn{5}{c}{\textbf{Beauty}} \\
\textbf{token(s)} & & \textbf{N@2} & \textbf{N@5} & \textbf{H@3} & \textbf{CP@3} & \textbf{CD} \\
\hline 
\multirow{8}{*}{\rotatebox{90}{\textbf{\textless category\textgreater}}} & SASRec & 0.2469 & 0.2867 & 0.1536 & 0.2512 & 3.4857 \\
& BERT4Rec & 0.2561 & 0.2721 & 0.1458 & 0.2271 & 3.6612 \\
& GRU4Rec & 0.1839 & 0.2063 & 0.1016  & 0.2178 & 4.1269 \\
\cline{2-7}
& LLaRA & 0.4089 & \underline{0.4210} & \underline{0.2172} & \underline{0.3565} & 2.4193 \\
& LlamaRec & 0.2494 & 0.2924 & 0.1321 & 0.2527 & 3.3724 \\
& TALLRec & \underline{0.4937} & 0.4172 & 0.2092 & 0.3264 & \underline{2.3629} \\
\cline{2-7}
& \textbf{{\name}} & \textbf{0.6175} & \textbf{0.4999} & \textbf{0.2350} & \textbf{0.4786} & \textbf{2.0071} \\
\hline
\end{tabular}
\end{table}

\subsubsection{Baseline Methods.}

We compare our method with two classes of baselines: (1) Sequential recommender systems, including: SASRec \cite{kang2018sasrec}, BERT4Rec \cite{sun2019bert4recsequentialrecommendationbidirectional}, GRU4Rec \cite{gru4rec}; (2) LLM-based recommender systems, including: LLaRA \cite{liao2024llaralargelanguagerecommendationassistant}, LlamaRec \cite{yue2023llamarectwostagerecommendationusing}, TALLRec \cite{Bao_2023}.  Our baseline evaluation focuses on ranking performance. For two-stage re-rankers, we provide them with identical candidate sets generated by our SASRec model. This standardized approach isolates the evaluation to purely compare re-ranking capabilities. Detailed implementation and introduction of baselines are in Appendix B.

\subsection{Main results. (RQ1) (RQ3)}

Table~\ref{COREC-results_1} and Table~\ref{COREC-results_2} present the experimental results across three datasets. Across various control token combinations, {\name} demonstrates a significant improvement on average recommendation performance, ranging from 17.3\% to 40.4\%. Our model exhibits superior performance across control effectiveness metrics as well, achieving an average improvement of 33.4\% in control precision over the second-best performing model, while reducing control depth by 12.5\%. These results demonstrate that {\name} significantly enhances the controllability of recommendation outcomes, ensuring that generated recommendations align closely with the specified control tokens than existing approaches.

\subsection{Further Analysis}

\subsubsection{Model Effectiveness. (RQ1)}
\begin{figure}[ht]
    \centering
    \subfigure[]{
        \includegraphics[width=0.2\textwidth]{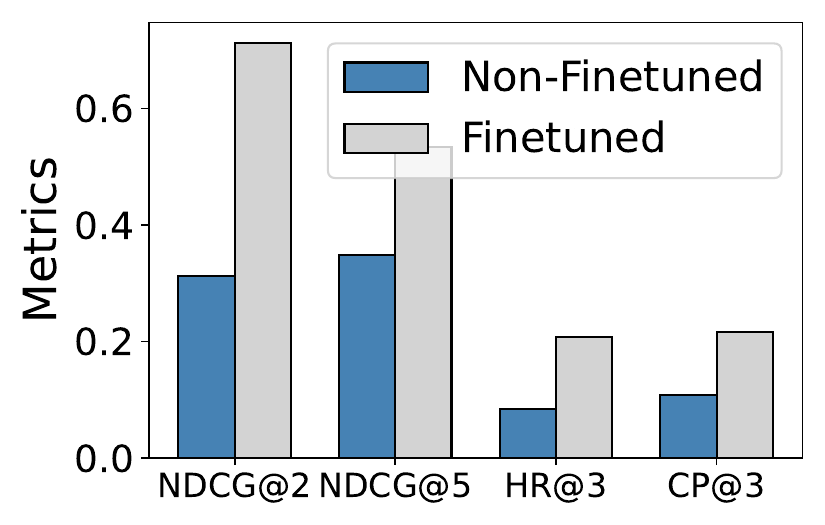}
        \label{fig:a}
    }
    \subfigure[]{
        \includegraphics[width=0.2\textwidth]{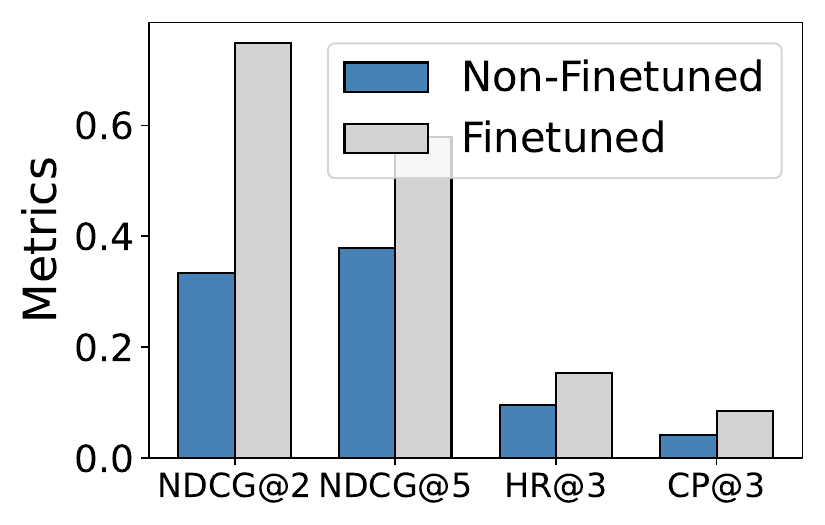}
        \label{fig:b}
    }
    \caption{Zero-shot vs {\name} comparison on Home dataset. (a) Two-token scheme; (b) Three-token scheme.}
    \label{non-vs-fine-tune}
\end{figure}

We evaluated our fine-tuned model against zero-shot inference, where recommendations were generated directly without fine-tuning. As shown in Figure \ref{non-vs-fine-tune}, the zero-shot approach showed notably lower performance compared to our fine-tuned model, despite receiving identical candidate items from the retrieval stage and the same control token definitions in the prompt. This outcome indicates that defining control tokens in the prompt is inadequate for recommendation control. Instead, incorporating learnable embeddings for these tokens enables the model to disentangle user intent into specific attribute requirements, leading to more targeted recommendations.

\begin{figure}[ht]
    \centering
    \subfigure[]{
        \includegraphics[width=0.22\textwidth]{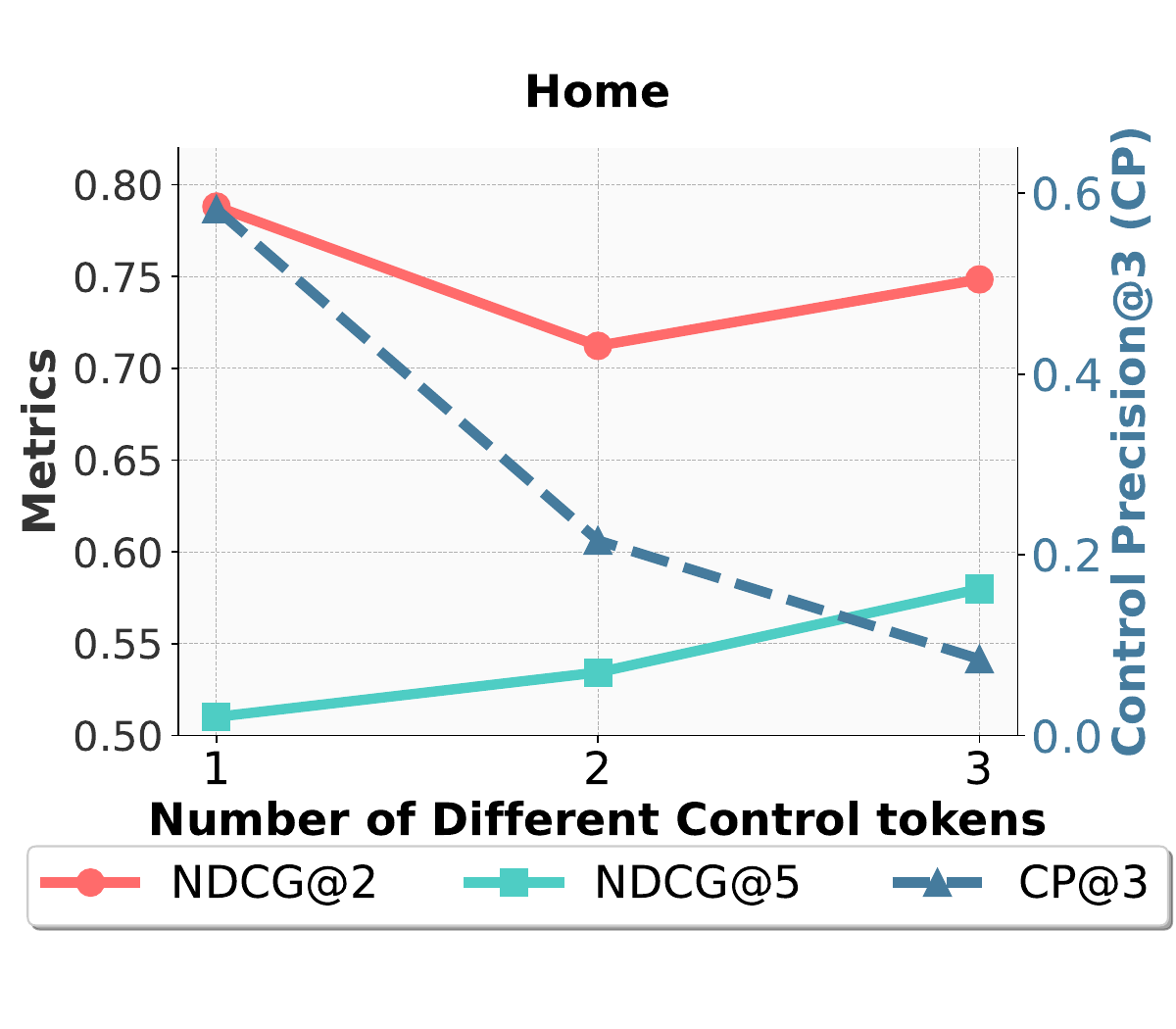}
        \label{fig:a}
    }
    \subfigure[]{
        \includegraphics[width=0.22\textwidth]{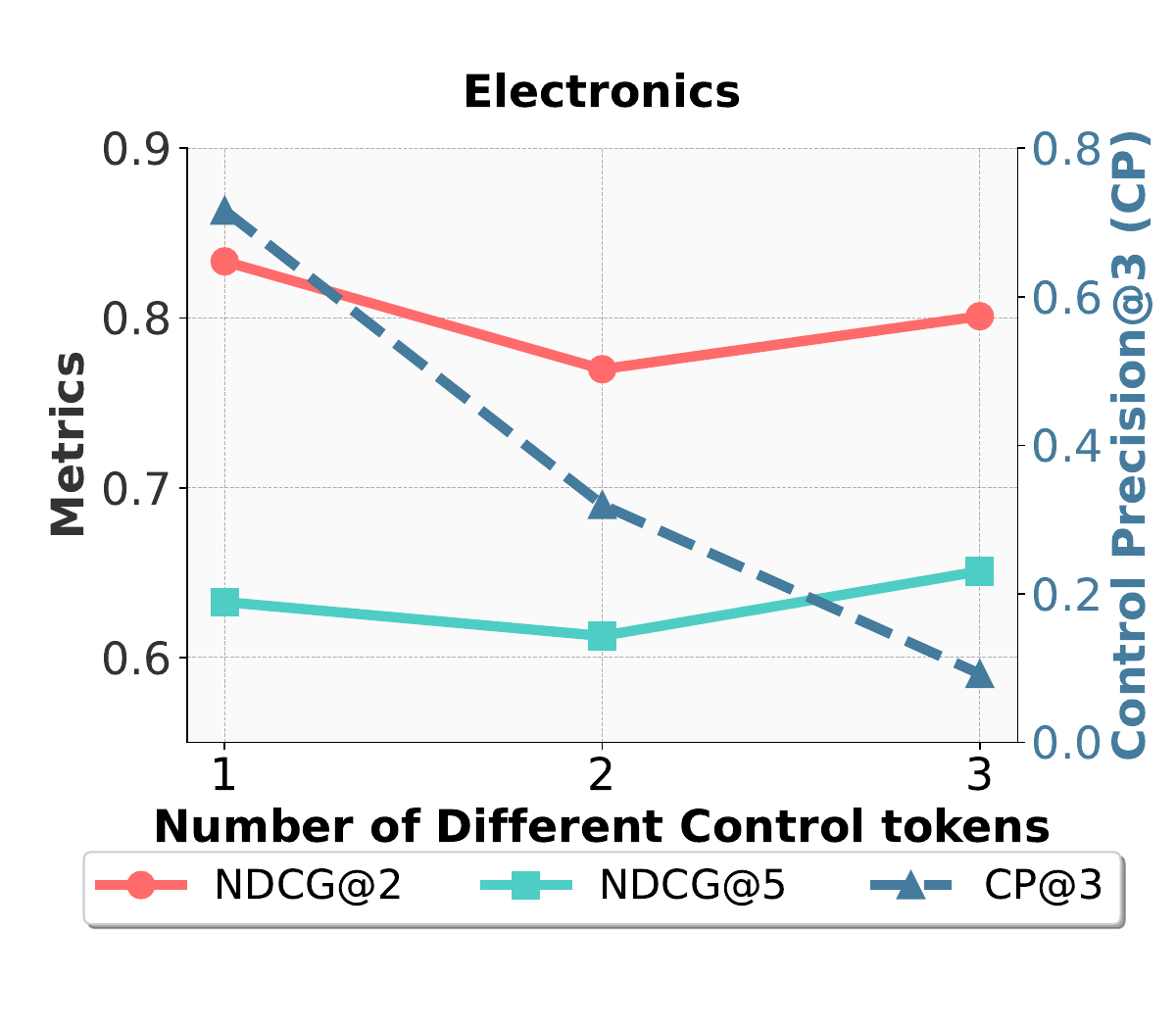}
        \label{fig:b}
    }
    \caption{{\name} Performance by Control Token Count.} 
    \label{impact-tokens}
\end{figure}

\subsubsection{Impact of Control Token Count. (RQ2)}
As shown in Figure~\ref{impact-tokens}, our model demonstrates distinct behavioral patterns as control token count increases. The transition from \textbf{single-token} to \textbf{two-token} control reveals a precision-relevance trade-off: while NDCG remains relatively stable, control precision (CP) decreases markedly. This occurs because fewer candidates within the entire candidate set can satisfy the stricter multi-token requirements under the maximum relevance threshold, naturally reducing CP. The enhanced constraints impose a minor cost on NDCG as well. 

This trend reverses with the \textbf{three-token} scheme, where NDCG begins to improve as the model adapts its ranking strategy under tighter constraints, though CP continues its expected decline. Despite this natural decrease in CP with increased control complexity, our approach maintains substantially higher control precision than baseline methods across all token configurations, as shown in Table~\ref{COREC-results_1}.

\begin{figure}[ht]
    \centering
    \subfigure[]{
        \includegraphics[width=0.22444\textwidth]{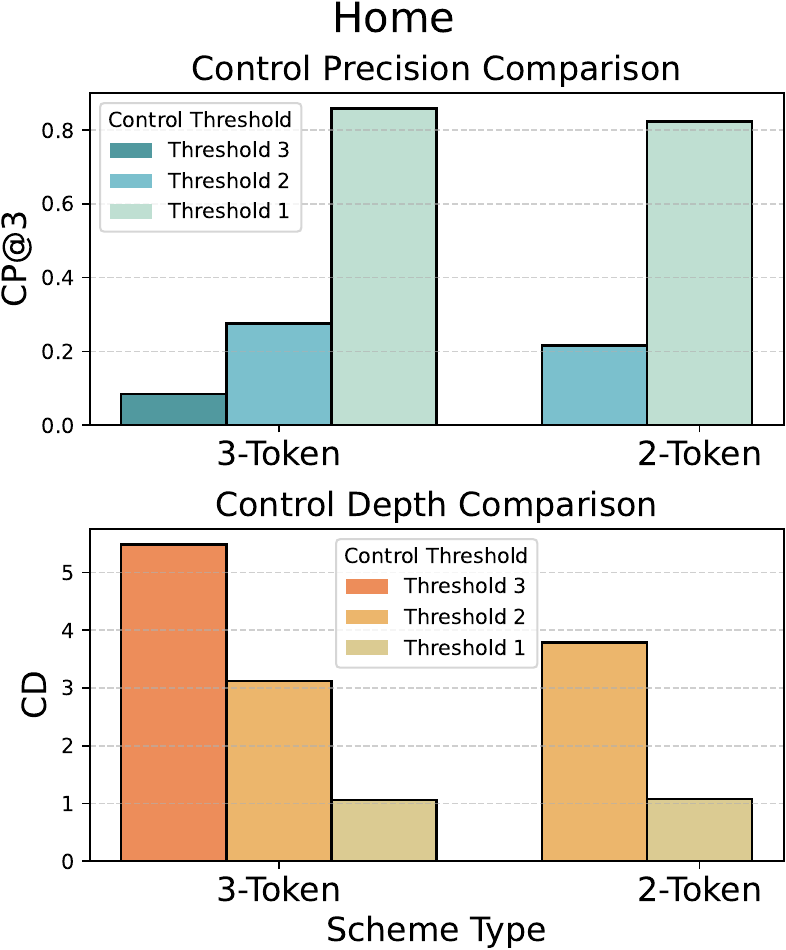}
        \label{fig:a}
    }
    \subfigure[]{
        \includegraphics[width=0.22444\textwidth]{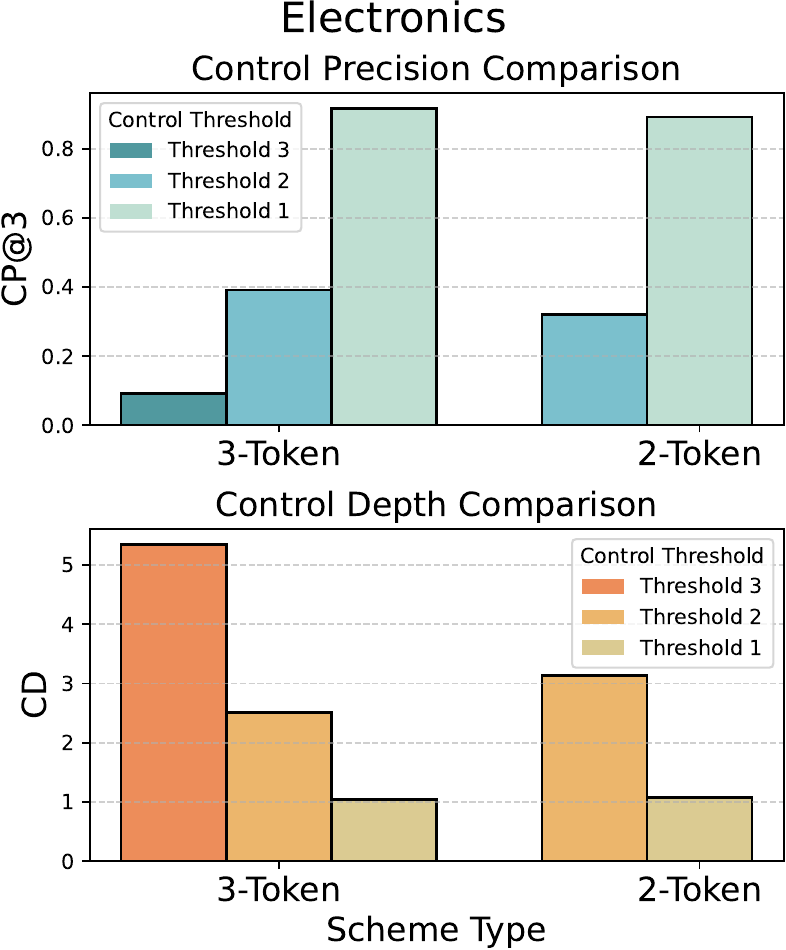}
        \label{fig:b}
    }
    \caption{Controllability performance of {\name} under descending control thresholds. }
    \label{threshold-comparison}
\end{figure}

\begin{figure}[ht]
    \centering
    \subfigure[]{
        \includegraphics[width=0.22444\textwidth]{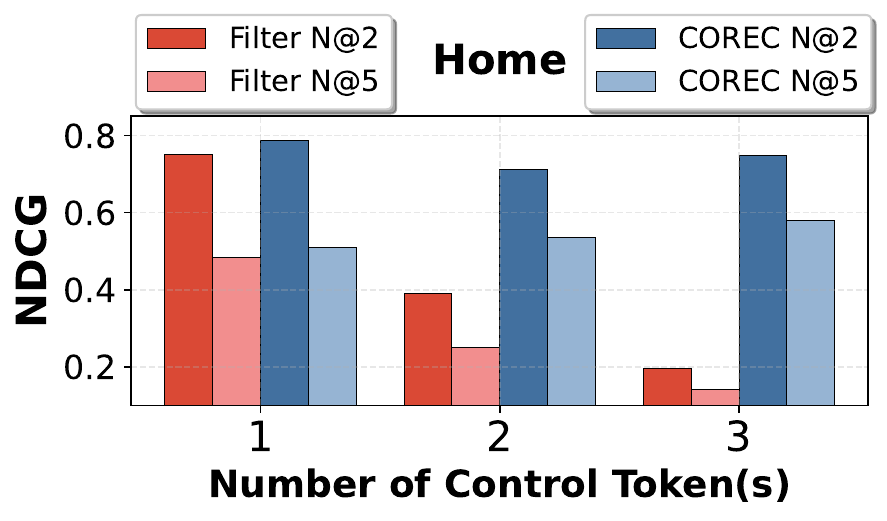}
        \label{fig:a}
    }
    \subfigure[]{
        \includegraphics[width=0.22444\textwidth]{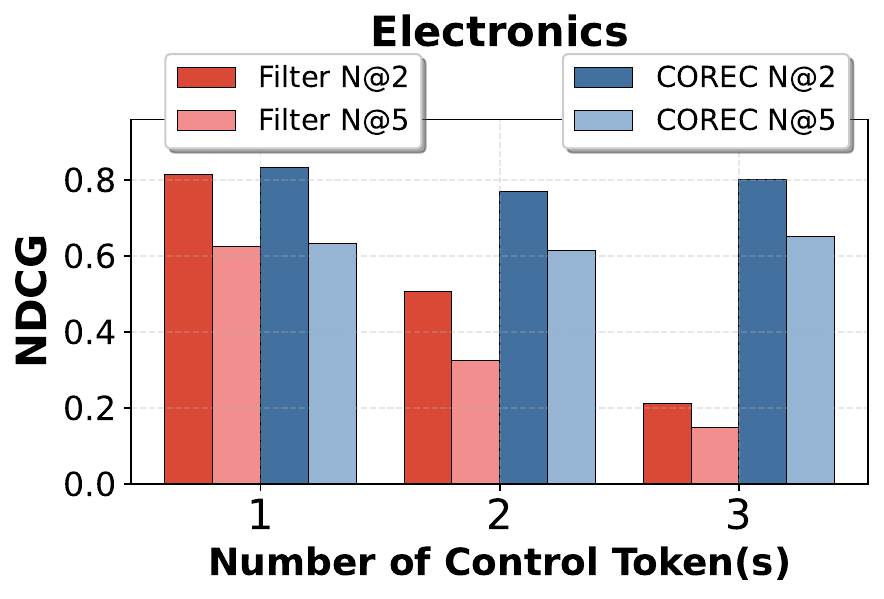}
        \label{fig:b}
    }
    \caption{Performance comparison between hard-filter method and ours across varying numbers of control tokens. }
    \label{filter-comparison}
\end{figure}

\subsubsection{Impact of Control Threshold. (RQ3)}

Table~\ref{COREC-results_1} and Table~\ref{COREC-results_2} present our initial results where control precision (CP) and control depth (CD) were assessed against the maximum control threshold. Building upon the previous section, we observe that CP declines as control token count increases. Next, we investigate whether this decline results from the stringent maximum control thresholds.

To examine this hypothesis, we systematically adjust the control threshold for multi-token schemes and monitor both CP and CD. As illustrated in Figure~\ref{threshold-comparison}, reducing the threshold substantially improves control performance for both 2-token and 3-token schemes, with the 3-token configuration consistently outperforming the 2-token approach at equivalent thresholds. This demonstrates that even if the measured metrics decrease as the number of control tokens rises, our model maintains effective controllability over recommendation outcomes according to specified attributes.

\subsubsection{Filtering Comparison. (RQ4)}
To validate our model's capacity for balancing control and recommendation quality, we compare {\name} against a hard-filtering baseline that strictly excludes all items not matching the specified control tokens, where control effectiveness metrics naturally reach their maximum values. This baseline employs SASRec as its backbone, consistent with our retrieval stage method. 

We evaluate both approaches under multi-token control scenarios, applying the strictest control threshold as in Table~\ref{COREC-results_1}. As illustrated in Figure~\ref{filter-comparison}, the hard-filtering approach exhibits declining NDCG performance, with this decline becoming more pronounced as control token count increases. These results confirm our hypothesis that applying multiple constraints to finite candidate pools inherently compromises ranking quality, thereby validating our method's superiority in maintaining the precision-relevance balance.

\section{Conclusion}
In this work, we introduce {\name}, a novel framework that enhances the controllability of LLM-based recommendations. The core of our approach is the implementation of control tokens, a technique we propose to encode attribute-level user requirements. We demonstrate how tokens provide flexible recommendation guidance. Extensive experiments show that {\name} successfully enforces these explicit user constraints while maintaining competitive performance on recommendation quality. This research provides a robust solution for building more trustworthy and interactive recommender systems, and future work could explore extending this control mechanism to more complex, multi-attribute user queries or dynamic, session-based constraints.

\bibliography{aaai2026}

\clearpage
\section{A. Supplementary Related Works}

\subsubsection{A.1 LLMs as Generative Recommenders\\}
Generative approaches have emerged as a prominent perspective in LLM-based recommendation systems \cite{wu2023surveyLLM4Rec, li2024largelanguagemodelsgenerative,Xu_2024}. To make LLM-based recommendations both reliable and efficient, existing controllable approaches primarily focus on mitigating hallucination and output regularization, typically employing two kinds of techniques: reinforcement learning and sequence tokenization.

Primarily, supervised or reinforcement learning is used during fine-tuning to regularize the recommendation task \cite{ you2025textr2texteclargerecommendermodels,lu-etal-2024-aligning}. Other prompt-based methods adopt a fill-in-the-blank paradigm \cite{chen-etal-2025-DLCRec, lu-etal-2024-aligning, zhang2025slowthinkingsequentialrecommendation} or re-ranking to enforce instruction execution. These methods primarily focus on standardizing the recommendation logic chain to adapt to different recommendation scenarios, aiming to maximize the utilization of LLMs' capabilities in generating textual information. However, they exhibit limitations in providing the flexibility to accommodate dynamic, user-specified control requirements at the item attribute level.

Simultaneously, sequence tokenization is widely adopted for grounding LLM outputs in real-world items, therefore avoiding LLM hallucination. These methods tokenize both users and items as discrete token sequences, then compute similarity between user behavior and generated results \cite{Rajput-etal-tiger,qu2024tokenreclearningtokenizeid,wang2024learnableitemtokenizationgenerative,liu2025generativerecommenderendtoendlearnable, liao2024llaralargelanguagerecommendationassistant} for efficient item indexing. However, semantic compression into tokens causes LLMs to lose specific item information compared to full-length prompts, rendering these methods inadequate for fine-grained recommendation control at the attribute level.

\section{B. Experiment}
\subsection{B.1 Dataset Pre-processing}
\begin{figure}[ht]
    \centering
    \includegraphics[width=0.23\textwidth]{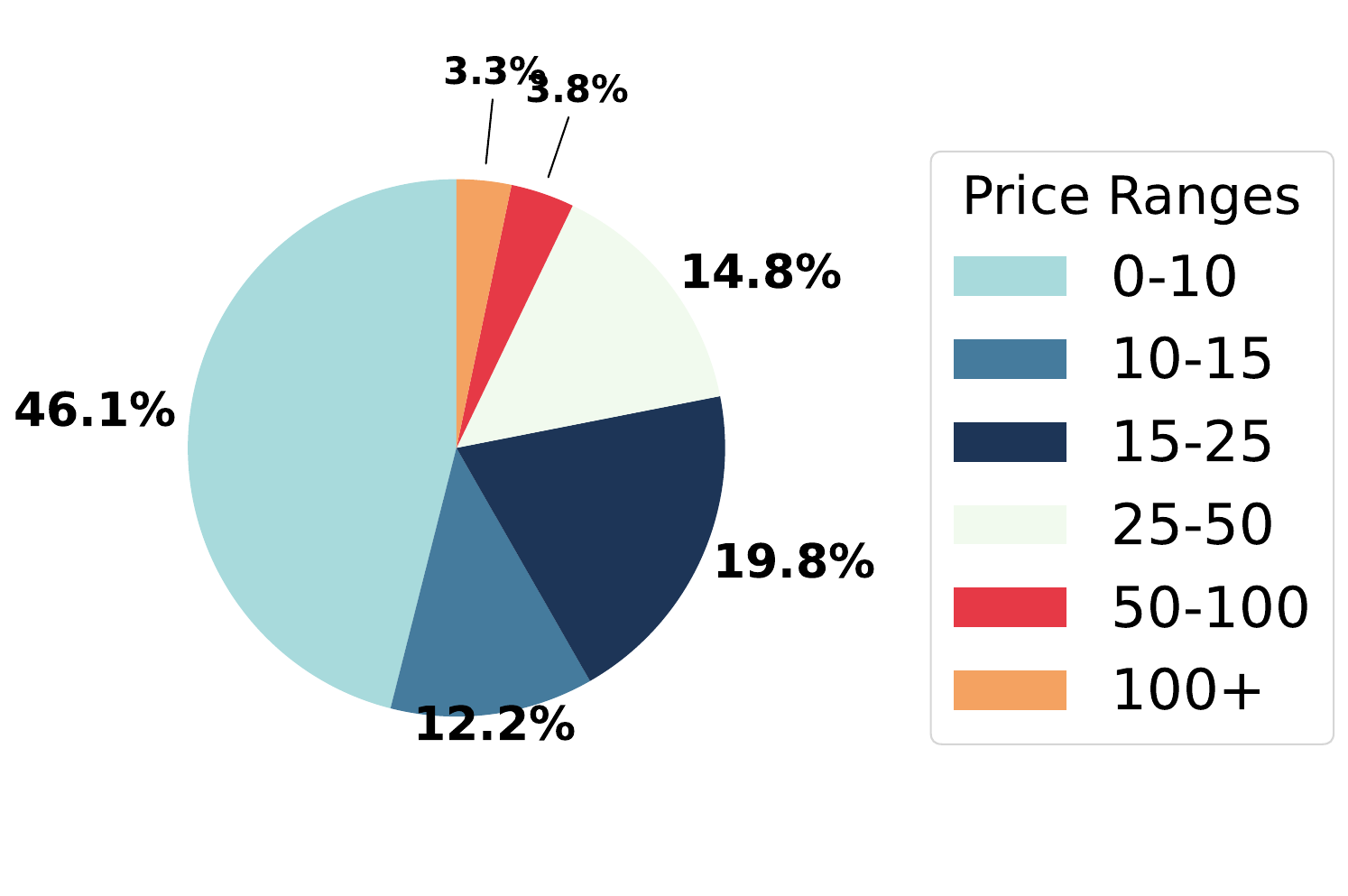}
    \hfill
    \includegraphics[width=0.23\textwidth]{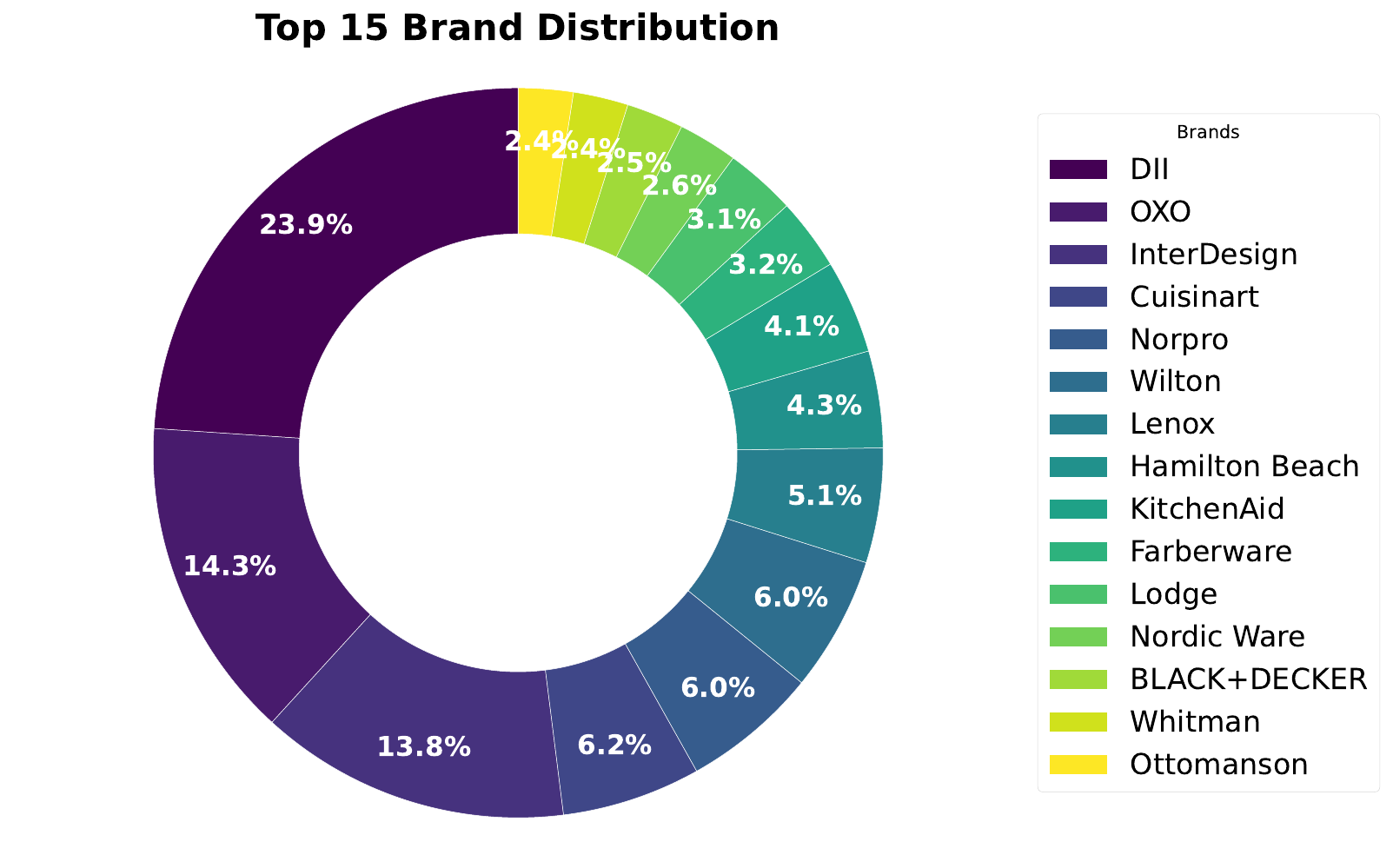}
    \caption{Token distributions of price and brand.}
    \label{token_dist}
\end{figure}

During the data pre-processing stage, we first classify any rating above 3 as positive feedback. To satisfy the control token evaluation criteria, we pre-process the data by filtering out entries with missing values for the relevant control attributes—specifically price, rank, and brand in the Home and Electronics datasets, and category in the Beauty dataset. User actions are considered valid interactions only when both conditions are satisfied. We then employ timestamps to establish the chronological order of these actions. To ensure all training sequences are sufficiently informative, we pre-filter users based on their interaction history. For the Home and Kitchen and Electronics datasets, we randomly select 3,000 users, each with a minimum of 30 interactions. To unify sequence lengths for users with extensive histories, we truncate their records to retain only their 50 most recent interactions. Given the greater sparsity of the All Beauty dataset, we adjust our filtering criteria. For this dataset, we include all users with at least 6 interactions. To ensure a robust evaluation, its validation and test sets are constructed exclusively from users with a minimum of 8 interactions. We employ a sliding window approach to construct data for both the item retrieval stage and LLM fine-tuning stage, with a unified window size of 6, where the final item at position 6 serves as the prediction target. We present a comprehensive distribution analysis of various control token types, as shown in Figures~\ref{token_dist}.

\subsection{B.2 Implementation Details}
Regarding the implementation details of {\name}, we fine-tune LLaMA-3-8B (with a hidden dimension of 4,096) using LoRA \cite{hu2021loralowrankadaptationlarge} for parameter-efficient training. The LoRA configuration employs a rank of 8, an alpha scaling factor of 16, and a dropout rate of 0.05. Training is conducted for 3 epochs using DeepSpeed \cite{rajbhandari2022deepspeedmoeadvancingmixtureofexpertsinference} ZeRO Stage 2 for distributed optimization. We utilize a per-device batch size of 2 with gradient accumulation steps of 1, and input sequences are truncated to a maximum length of 2,048 tokens. The best results introduced in our paper's experiment section are all reported based on the highest Hit Rate metric on the validation set. 

\subsection{B.3 Compared Methods}
\subsubsection{B.3.1 Sequential Recommendation Methods\\}
For sequential recommendation, model implementation typically requires only the user and item ID information along with timestamps. Traditional sequential recommendation methods predominantly rely on encoding user interaction behaviors through various approaches to capture and infer both long-term and short-term user preferences from historical interaction sequences. 

To implement these baselines, we searched the learning rate from the set \{1e-2, 1e-3, 5e-3, 1e-4\}.  The embedding dimension was searched over \{32, 64, 128\}, and we experimented with a batch size of 128 and a dropout rate of 0.2. 

\begin{itemize}
    \item \textbf{SASRec} \cite{kang2018sasrec}
    employs self-attention mechanisms to model users' historical behavioral patterns and extract valuable sequential information. The model computes inner products between the learned representations and all item embeddings, then ranks and filters the results based on relevance scores to generate Top \(K\) recommendations.
    \item \textbf{GRU4Rec} \cite{hidasi2016sessionbasedrecommendationsrecurrentneural} introduces RNN-based sequential recommendation, treating user interactions as temporal sequences to capture evolving preferences and predict next items in variable-length sessions.
    
    \item \textbf{BERT4Rec} \cite{sun2019bert4recsequentialrecommendationbidirectional} uses deep bidirectional self-attention to capture dependencies between items in user behavior sequences for next item prediction. Unlike traditional unidirectional models, it leverages all items both before and after a given item as context, enabling more comprehensive sequence representations.
\end{itemize}
\subsubsection{B.3.2 LLM-based Sequential Recommendation Methods\\}
We select several representative LLM-based sequential recommendation models as baselines. For fair comparison, we follow the original experimental configurations and hyperparameter settings as specified in their respective source code implementations. Specifically, we reproduce TALLRec and LlamaRec using Llama3-8B. For LLaRA, we utilize Llama2-7B following the authors' original framework, as this represents the optimal configuration validated in their code implementation. As to inference stage, LLARA generates single-item titles token by token, but re-ranking is required for a fair comparison. We compute the conditional probability of generating each candidate’s title and rank by using the model’s next-token probability distribution, and then sort the candidates accordingly. LlamaRec inherently follows this approach. TALLRec performs a binary judgment for each candidate, and we rank them based on the predicted probabilities. We adopt this reproduction setting, which is reasonable as it aligns with our method as well.

\begin{itemize}
    \item \textbf{LlamaRec} \cite{yue2023llamarectwostagerecommendationusing}
    employs a two-stage re-ranking approach for candidate item ordering, where the LLM utilizes a verbalizer to map token logits from the LLM output head to probability scores for each candidate. This enables obtaining ranking scores for all candidates in a single forward pass, significantly improving both training and inference efficiency.
    \item \textbf{TALLRec} \cite{Bao_2023}
    uses two-stage instruction tuning that decomposes user interactions into binary responses for few-shot training, aligning general LLMs with recommendation tasks.
    \item \textbf{LLaRA}  \cite{liao2024llaralargelanguagerecommendationassistant}
    utilizes a hybrid representation that combines text tokens with behavior tokens from traditional recommenders, using prompt tuning to align user behavior patterns with LLM for improved sequential recommendation performance.
\end{itemize}
\subsection{B.4 {\name} Further Studies}
\subsubsection{B.4.1 Ablation Study on Control Token Effectiveness\\}
\begin{figure}[ht]
    \centering
    \includegraphics[width=0.223\textwidth]{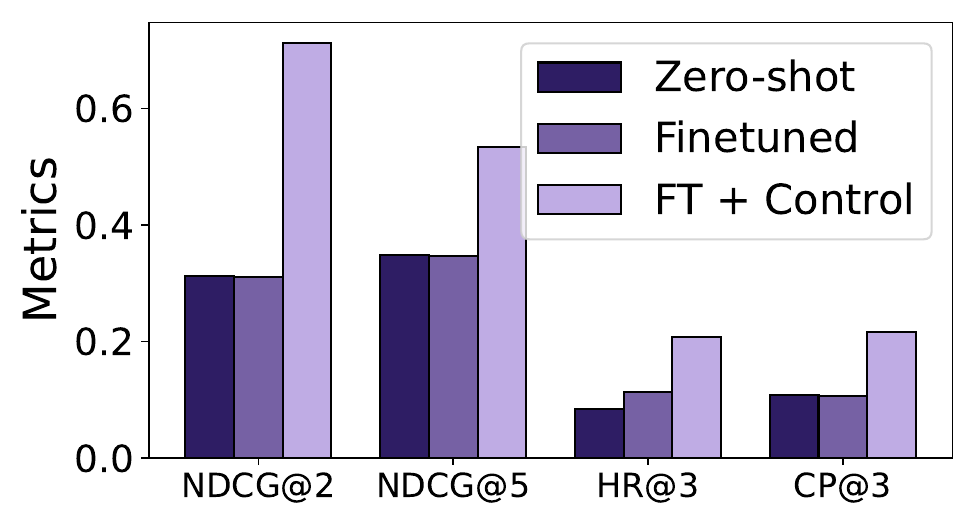}
    \hfill
    \includegraphics[width=0.223\textwidth]{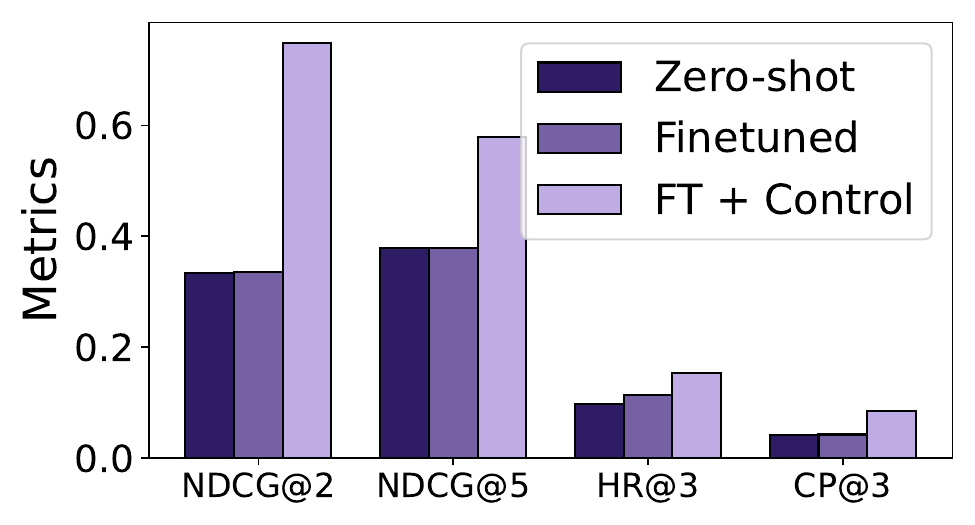}
    \caption{2 and 3 token scheme.}
    \label{vs result}
\end{figure}

To further assess the effectiveness of our approach, we perform a comparative analysis across multiple configurations: zero-shot inference, full fine-tuning without control tokens, and our proposed control token-based fine-tuning method. Figure~\ref{vs result} illustrates the experimental results. Our findings reveal that without the incorporation of control tokens, the model exhibits limited improvement even after fine-tuning, particularly demonstrating minimal gains in control effectiveness.

\section{C. Prompt Templates and Examples}
\subsection{C.1 GPT-based Data Augmentation Prompt}
See Figure~\ref{prompt1} below.

\subsection{C.2 {\name} Prompt Example}
See Figure~\ref{prompt2} below.

\begin{figure*}[t]
\centering
\includegraphics[width=\textwidth]{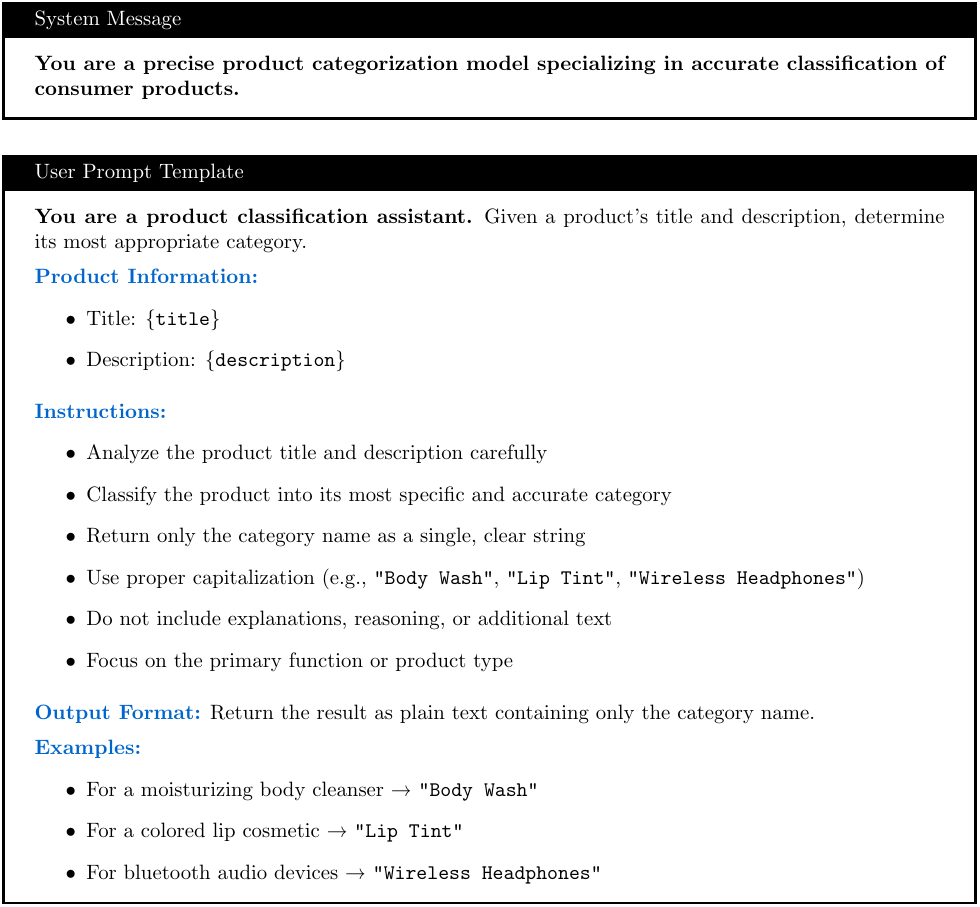}
\caption{Data Augmentation Template}
\label{prompt1}
\end{figure*}

\begin{figure*}[t]
\centering
\includegraphics[width=\textwidth]{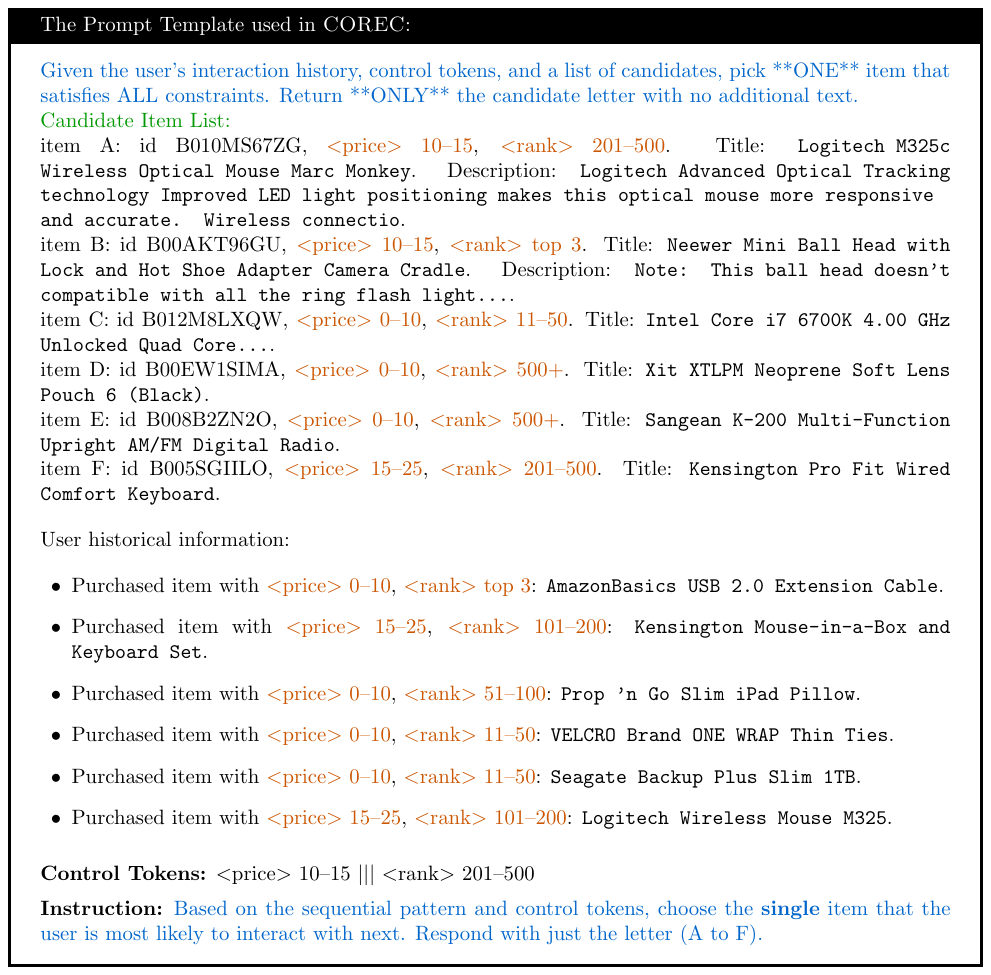}
\caption{{\name} Prompt Example}
\label{prompt2}
\end{figure*}
\end{document}